\documentclass[aps,12pt,eqsecnum,nofootinbib,preprintnumbers,
               tightenlines,dvips,showpacs]{revtex4}
\usepackage{graphicx,bm,amsmath,amssymb}

\begin{document}
\newcommand{\etal}{{\it et al.,\;}}

\preprint{INT PUB 07-27}

\title{Renormalization group flow of quartic perturbations in graphene:
    Strong coupling and large-$N$ limits}
\author{Joaqu\'{\i}n E. Drut and Dam Thanh Son}
\affiliation{Institute for Nuclear Theory, University of Washington, 
Seattle, WA 98195--1550, USA}

\begin{abstract}
We explore the renormalization group flow of quartic perturbations in
the low-enegy theory of graphene, in the strong Coulomb coupling and
large-$N$ limits, where $N$ is the number of fermion flavors. We
compute the anomalous dimensions of the quartic couplings $u$ up to
leading order in $1/N$ and find both relevant and irrelevant
directions in the space of quartic couplings. We discuss possible
phase diagrams and relevance for the physics of graphene.
\end{abstract}

\date{\today}

\pacs{73.63.Bd, 05.10.Cc}

\maketitle

\section{Introduction}

Graphene, a single layer of graphite, has lately attracted increasing
attention, especially since its experimental realization
\cite{GeimNovoselov}. This material presents a number of unusual
electronic features, such as a Landau level structure that gives rise
to half-integer quantum Hall effect. In general, those properties can
be traced back to the low-energy spectrum, governed by two-component
massless fermions that come in four flavors: two due to
electronic spin and two from degenerate Dirac points in the band
structure \cite{Semenoff}. 
The quasi-relativistic electronic spectrum
is characterized by a velocity $v \simeq
c/300 \ll c$, where $c$ is the speed of light.  However, the Coulomb
interaction, which is instantaneous for all practical purposes,
breaks this emergent Lorentz invariance.  In particular, the electron
velocity becomes scale-dependent~\cite{Gonzalez:1993uz}.

It was noted recently \cite{Son} that a
generalizaton of the low-energy theory that describes graphene
possesses a quantum critical point.
Reference~\onlinecite{Son} considered a model of $N$ species of $2+1$
dimensional two-component massless Dirac fermions interacting through
a 3D instantaneous Coulomb interaction with a coupling constant $g$.
It was then shown, by
analyzing the renormalization group (RG) flow of the fermion velocity,
that for sufficiently large $N$, the strongly coupled limit 
$g^2N/v\to\infty$ is
a quantum critical point, characterized by a dynamic critical exponent
$z=1 - 8/(\pi^2 N) + O(1/N^2)$.  It was also argued that real graphene
is close to this critical point for a large momentum window.

In the present work we further explore the quantum critical properties
of this theory by studying the RG flow of various quartic
perturbations of the Thirring \cite{Thirring} and Gross-Neveu
\cite{Gross-Neveu} varieties.  These four-fermion interactions are 
irrelevant at weak coupling, but obtain nontrivial anomalous dimensions
of order $1/N$ at the strong coupling fixed point.  If a four-fermi coupling
becomes relevant below some $N$, one may expect dynamical gap generation,
in which case the system becomes an insulator,
or that the system would flow to another fixed point.

The question of whether the Coulomb interaction makes graphene an
insulator has been considered previously.  In
Ref.~\onlinecite{Gorbar:2002iw} it was found from solving the gap
equation with a screened Coulomb interaction that, for $N<8/\pi\approx2.55$,
a gap opens at sufficiently large coupling.  For $N>8/\pi$ the system
remain gapless at all couplings.  This was confirmed in
Ref.~\onlinecite{Leal:2003sg} using a similar approach.  However, as
the approximations employed in these works are uncontrolled, one would
like to explore alternative approaches to this problem.  The RG
analysis in this paper is one of them. (Instability toward ferromagnetism
has also been considered~\cite{Peres}.)

We implement a Wilson-Fisher RG transformation and find the anomalous
dimensions of the various couplings. Throughout the paper the Coulomb
parameter $\lambda = g^2N/32v$ is kept finite and the limit $\lambda
\rightarrow \infty$ is discussed at the end. In this limit we identify
both relevant and irrelevant directions in the parameter space of the
quartic couplings and discuss possible scenarios, including gap
generation and the the flow to a new non-gaussian fixed point. At the
end of our results section we address the relevance of our findings
for real graphene.

The paper is organized as follows: the model and Feynman rules are
presented in section II, the details of the calculation are outlined
in section III, the results are discussed in section IV and the conclusions 
presented in section V.

\section{Model and Feynman rules}

Consider a model of $N/2$ flavors of four-component relativistic
fermions (or $N$ flavors of two-component fermions) $\psi_a$
interacting through an instantaneous Coulomb interaction and through
generic U($N/2$)-symmetric quartic interactions.
The corresponding Euclidean action is
\begin{equation}\label{SE}
\begin{split}
  S_E &= -\int\! dt\, d^2x\, 
  \left( \bar \psi_a \gamma^0 \partial_0 \psi_a 
  + v \bar \psi_a \gamma^i \partial_i \psi_a 
  + i A_0 \bar\psi_a \gamma^0 \psi_a \right ) 
  + \frac{1}{2g^2}\int\!dt\, d^3x  (\partial_i A_0)^2 \\
  &\quad + \frac1{2N} \int\!dt\, d^2x\, \left[ u_1 (\bar\psi_a \psi_a)^2
  + u_2 (\bar\psi_a \gamma_0\gamma_i\psi_a)^2
  + u_3 (\bar\psi_a \gamma_0 \gamma_5 \psi_a)^2 
  + u_4 (\bar\psi_a \gamma_0 \psi_a)^2 \right.\\
  &\qquad\qquad\qquad\qquad \left.
  + u_5 (\bar\psi_a \gamma_i \psi_a)^2 
  + u_6  (\bar\psi_a \gamma_5\psi_a)^2\right]
\end{split}
\end{equation}
where the $\gamma^\mu$, $\mu = 0,1,2$ (we shall also use Latin indices
for the spatial directions) are Dirac matrices satisfying a Euclidean
Clifford algebra $\{ \gamma^\mu, \gamma^\nu\} = 2\delta^{\mu \nu}$, 
and $\gamma^5=\gamma^0\gamma^1\gamma^2$. We
can choose, for example, a representation of this algebra in which
\begin{equation}
 \gamma^0 = \left(\begin{array}{cc} 
    \sigma^3 & 0 \\ 0 & -\sigma^3 \end{array}\right),
 \qquad 
 \gamma^i = \left(\begin{array}{cc} 
    \sigma^i & 0 \\ 0 & -\sigma^i \end{array}\right),
\end{equation} 
where $\sigma^i$ are the Pauli matrices. In the real world $N/2=2$,
corresponding to two spin polarizations of the electrons.  The two
two-component spinors near the two valleys are combined into a
four-component spinor.  The action~(\ref{SE}) is invariant under spin
rotations, but is not invariant under rotations in the valley 
space.\footnote{The two degenerate Dirac points in the band structure 
of graphene are sometimes referred to as valleys. Thus, rotations in 
valley space correspond to operations that mix the upper and lower 
components of the four-component Dirac fermions of our theory.} 
Notice that the four-fermi terms are rotationally invariant, but not
Lorentz invariant: the latter is broken by the Coulomb interaction.
Apart from the four-fermi terms written in~(\ref{SE}), one can also introduce 
an independent set equal to $\gamma^3$ times the vertices already included, 
where $\gamma^3$ is linearly independent from the other $\gamma$-matrices 
and $\{\gamma^3,\gamma^\mu\} = 0$. One possible choice is 
\begin{equation}
 \gamma^3 = \left(\begin{array}{cc} 
    0 & \openone \\ \openone & 0
    \end{array}\right).
\end{equation}
Such matrices, together with the ones already considered, form a 
complete basis for the algebra of $4\times4$ Dirac matrices. We restrict 
for the moment to the action (\ref{SE}), but we shall comment on these 
extra vertices in our discussion section.

We shall use the large-$N$ limit to do calculations at large values of
the coupling $g$, so $N$ will remain arbitary (but large) until the
end. Notice that, unlike the fermionic degrees of freedom, the Coulomb
field $A_0$ lives in $3+1$ dimensions, which is reflected in its
kinetic term. In the strong coupling $g \to \infty$ limit this term
disappears and the $A_0$ propagator is dominated by quantum
corrections coming from the fermion loops, as we shall see below.

\begin{figure}[h]
\includegraphics[width=8cm]{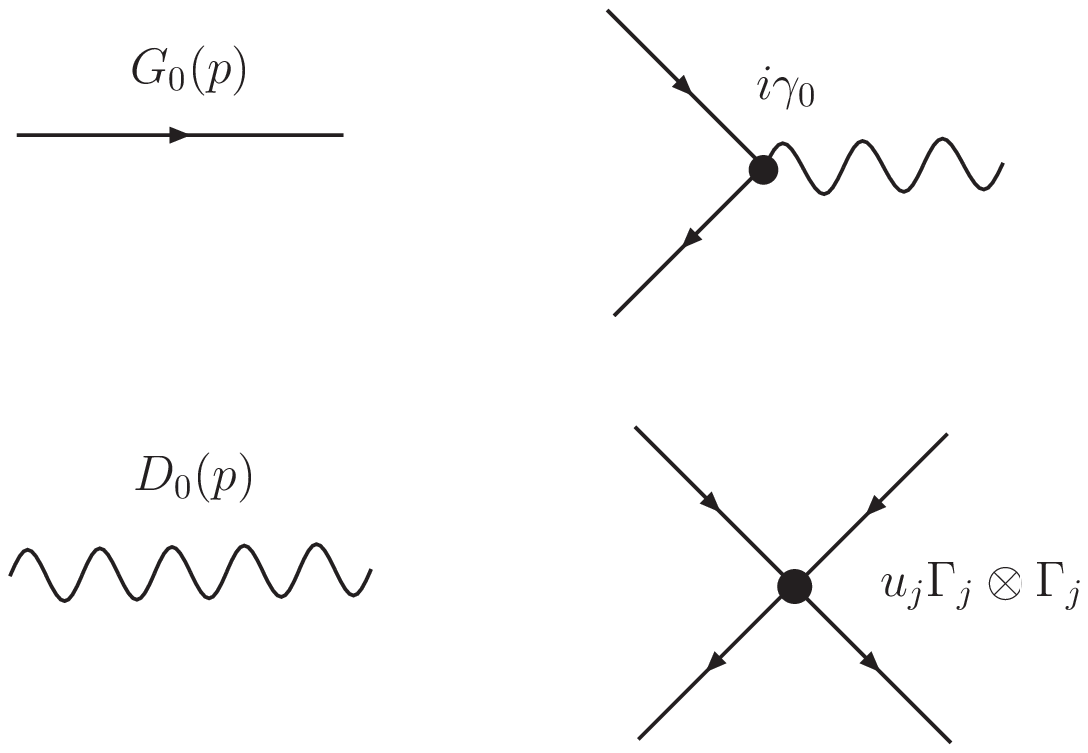}
\caption{\label{fig:FeynmanRules} Feynman rules (see text for details).}
\end{figure}

In $2+1$ dimensions the naive dimensions of the fields and couplings
are as follows: $[\psi] = m$, $[A_0] = m$, $[g] = m^0$, $[u_j] =
m^{-1}$.

The Feynman rules are as in Fig.~\ref{fig:FeynmanRules}, where the
fermion propagator is
\begin{equation}
G_0(p) = \frac{i \not p}{p^2},
\end{equation}
where $p = (p_0,{\vec p})$, ${\vec p}$ being a 2D momentum
vector. Here and in the rest of the paper, we use the notation ${\not
p} = \gamma^0 p_0 + v {\vec \gamma}\cdot {\vec p}$ and $p^2 = p_0^2 +
v^2 |{\vec p}|^2$.

The boson-fermion interaction vertex is $i\gamma_0$, and the quartic
vertices are $-\frac{u_j}{N} \Gamma_j \otimes \Gamma_j$, with
$\Gamma_j \in \{1, \gamma_0\gamma_i, \gamma_0\gamma_5, \gamma_0,
\gamma_i, \gamma_5 \}$.

\begin{figure}[h]
\includegraphics[width=16cm]{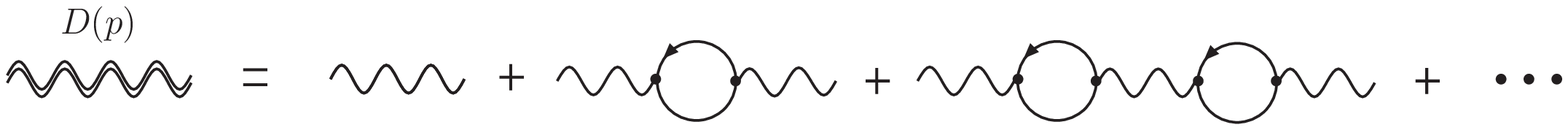}
\caption{\label{fig:BosonPropagator} Resummation of the one-loop 
self-energy contribution to the boson propagator.}
\end{figure}

Finally, the bare boson ($A_0$) propagator is
\begin{equation}
D_0(p) = \frac{g^2}{2|\vec{p}|}
\end{equation}
In the large $N$, finite $g^2N$ limit we must resum the 
fermion loop contributions to this
propagator (see Fig.~\ref{fig:BosonPropagator}).

\begin{figure}[h]
\includegraphics[width=6cm]{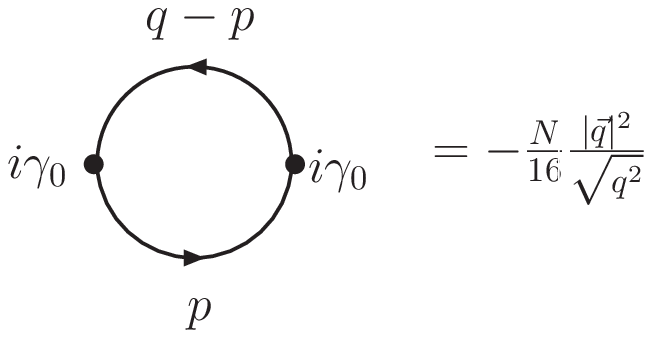}
\caption{\label{fig:FermionLoop} One-loop self-energy contribution 
to the boson propagator.}
\end{figure}
Such resummation (which is equivalent to the random phase
approximation) results in a dressed boson propagator that we shall use
in the rest of this work, as shown in Figs.~\ref{fig:BosonPropagator} and~\ref{fig:FermionLoop}:
\begin{equation}
  D(q) = \left( \frac{2 |\vec{q}|}{g^2} 
  + \frac{N}{16} \frac{|\vec{q}|^2}{\sqrt{q^2}} \right)^{-1} 
  = \frac{16}{N}\frac{\sqrt{q^2}}{|\vec{q}|^2}
  \left (1 + \frac{1}{\lambda} \frac{\sqrt{q^2}}{|\vec{q}|}\right)^{-1}
  = \frac{g^2}{2 |\vec{q}|}
  \left (1 + {\lambda} \frac{|\vec{q}|}{\sqrt{q^2}} \right)^{-1}
\end{equation}
where $\lambda = g^2 N/(32 v)$. 

\section{Calculation of the RG flow}

We implement a Wilsonian RG procedure whereby we integrate out modes
in the momentum shell $\Lambda_1 < p < \Lambda_0$ and then rescale the
coordinates as well as the fields.  The RG equation has the form
\begin{equation}
\frac{\partial {\bf u}(p)}{\partial \ln p} =
  (\openone-{\bf M}(\lambda)) {\bf u}(p)
\end{equation}
where $[{\bf u}]_i = u_i$ and ${\bf M}(\lambda)$ is a $6\times6$
matrix.  The $\openone$ in the right-hand side comes from the naive
dimension of $u_i$.  Furthermore 
\begin{equation}
  {\bf M}(\lambda) = {\bf M}_{v}(\lambda) + {\bf M}_{wf}(\lambda), 
\end{equation}
where ${\bf M}_v$ comes from the vertex corrections and ${\bf M}_{wf}$
comes from wavefunction renormalization.

\subsection{Vertex renormalization}

\begin{figure}[h]
\includegraphics[width=17cm]{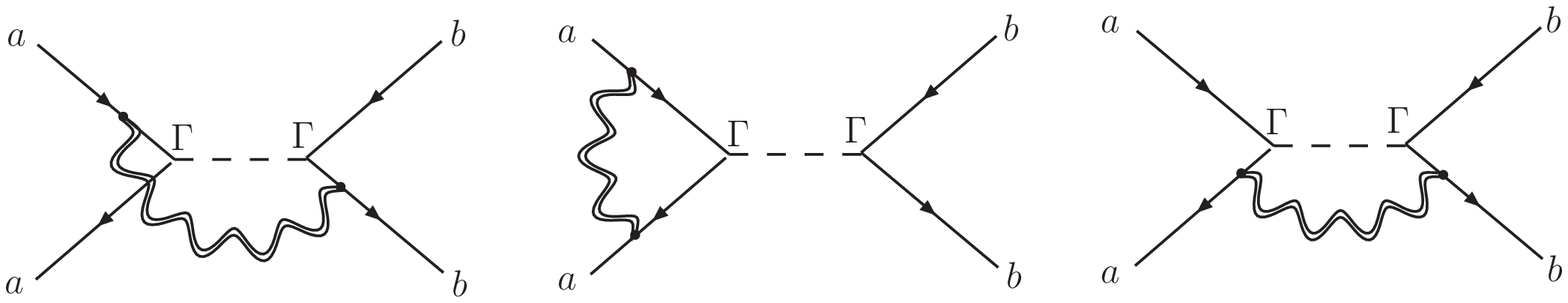}
\caption{\label{fig:diagrams} Contributions to the RG flow of $u_j$ 
coming from vertex renormalization. $a$ and $b$ are flavor indices and
$\Gamma$ is a generic four-Fermi vertex. The dashed line is intended
to show the difference between the loops in the various
diagrams. Exchange of dummy indices leads to the same diagrams and
results in an overall factor of 2.}
\end{figure}

To the first nontrivial order in $1/N$,
finding the RG flow of $u_j$ entails computing the diagrams in
Fig.~{\ref{fig:diagrams}} using the Feynman rules described in the
previous section. 
The loop integrals have the same form for all three diagrams, namely
\begin{eqnarray}
\label{eq:mainintegral}
\frac{16}{N}
\int_{\Lambda_1}^{\Lambda_0}\! \frac{d^3q}{(2\pi)^3} \frac{q^\mu q^\nu}{q^4} 
\frac{\lambda}{|\vec{q}|}
\left( 1 + \lambda \frac{|\vec{q}|}{\sqrt{q^2}}\right)^{-1} 
= \frac{4}{N\pi^2}\ln\left [\frac{\Lambda_0}{\Lambda_1} \right] 
&\times& I_0(\lambda)
\ \ \ \ \text{for $\mu = \nu = 0$}
\\ \nonumber
&\times& \frac{I_1(\lambda)}{2}
\ \ \ \ \text{for $\mu = \nu = i$}
\end{eqnarray}
and vanish for $\mu \neq \nu$. The integration region is spherically
symmetric. Furthermore, notice that the integrand is independent of
the azimuthal angle, and that the remaining radial and angular
integrals are decoupled because $|\vec{q}| = \sqrt{q^2}\sqrt{1-t^2}$,
where $t=\cos\theta$ and $\theta$ is the polar angle, making the
calculation straightforward. We have used the following definitions:
\begin{equation}
I_0(\lambda) = \int^1_{-1}dt
\frac{t^2}{\sqrt{1-t^2}}\frac{\lambda}{1+\lambda\sqrt{1-t^2}} = 
-2 + \frac{\pi}{\lambda} + \frac{2\sqrt{\lambda^2-1}}{\lambda} 
\ln\left[\lambda + \sqrt{\lambda^2-1} \right]
\end{equation}
\begin{equation}
I_1(\lambda) = \int^1_{-1}dt
\frac{\lambda\sqrt{1-t^2}}{1+\lambda\sqrt{1-t^2}} = 
2 - \frac{\pi}{\lambda} +
\frac{2}{\lambda \sqrt{\lambda^2 -1}}
\ln\left[ \lambda + \sqrt{\lambda^2-1} \right]
\end{equation}
In the large $\lambda$ limit, $\lambda \rightarrow \infty$,
\begin{equation}
I_0(\lambda) \rightarrow
-2 + 2 \ln\left[ \lambda + \sqrt{\lambda^2-1} \right]
\end{equation}
while
\begin{equation}
I_1(\lambda) \rightarrow 2 
\end{equation}
The divergence of $I_0(\lambda \rightarrow \infty)$ is associated the
the fact that the Coulomb interaction is unscreened at zero momentum
and nonzero frequency. As we shall see, however, all such divergences 
cancel out in the $\beta$ functions for $u_i$.

With the above identities at hand it is not very difficult to compute
the contributions of the diagrams in Fig.~\ref{fig:diagrams} to the
$\beta$ functions. The only significant step that remains to be discussed
is straightforward Dirac algebra to find the products of a small
number of $\gamma$ matrices (coming from the vertices and the fermion
propagators). For instance, for $\Gamma_4 = \gamma_0$, the calculation
of the diagram in the middle of Fig.~\ref{fig:diagrams} entails
computing
\begin{eqnarray}
\gamma_0\gamma_\mu\gamma_0\gamma_\nu\gamma_0
&=& \gamma_0 \ \ \ \ \ \ \ \text{for $\mu = \nu = 0$} \\
&=& -\gamma_0  \ \ \ \ \ \text{for $\mu = \nu = i$}
\end{eqnarray}
where we only need the case $\mu = \nu$ because the integral 
(\ref{eq:mainintegral}) is otherwise zero. The contribution of this diagram 
to the flow of $u_4$ is therefore
\begin{equation}
\Delta u_4 =  u_4 \frac{8}{N \pi^2}\ln\left [\frac{\Lambda_0}{\Lambda_1} 
\right](I_0 - I_1) 
\end{equation}
where we have included the factor of $2$ coming from an essentially
identical diagram where the boson propagator appears connecting the
fermion lines on the right instead of on the left. For this specific
vertex the first and third diagrams in Fig.~\ref{fig:diagrams} cancel
out exactly due to a sign difference in one of the fermion lines.
In this particular case only diagonal flow is generated, i.e. this
vertex does not contribute to the flow of the others. In general,
however, there is operator mixing: each vertex
contributes to its own flow but also to other vertices. 

Our result turns out to be
\begin{equation}\label{Mv}
{\bf M}_v(\lambda) =
\frac{8}{N\pi^2}
\left( \begin{array}{cccccc}
I_0 + I_1 & -2 I_1 & 0 & 0 & 0 & 0\\
-I_1 & I_0 & 0 & 0 & 0 & 0\\
0 & 0 & I_0 - I_1& 0 & 0 & 0\\
0 & 0 & 0 & I_0 - I_1 & 0 & 0\\
0 & 0 & 0 & 0 & I_0 & -I_1\\
0 & 0 & 0 & 0 & -2I_1 & I_0 + I_1
\end{array} \right)
\end{equation}
%

\subsection{Wavefunction renormalization}

Upon integration of the high-momentum degrees of freedom, new
coordinates and fields are defined, that are rescaled versions of the
original ones. Doing this explicitly results in the following action:
\begin{equation}
S_{\rm eff} = - \int\! dt'\,d^2x'\,
\left \{
Z_0^{-1} b_1^{-2} {\bar \psi}_a \gamma^0 \partial_0' \psi_a +
Z_1^{-1} b_1^{-1}b_0^{-1} {\bar \psi}_a \gamma^i \partial_i' \psi_a
+ ...\right \}
\end{equation}
where the subscript ``${\rm eff}$'' indicates that this action describes
modes with momenta $|p| < \Lambda_1 = b_1 \Lambda_0$. Here $t' = b_0
t$ $x' = b_1 x$, $0 < b_0,b_1 < 1$, and ... contains all the terms in
the action that are not quadratic in the fermion fields. We shall
renormalize the field as
\begin{equation}
\psi_a \rightarrow \psi_a' = Z_0^{-1/2} b_1^{-1} \psi_a
\end{equation}
and enforce the nonrenormalization of the fermion velocity by requiring
\begin{equation}
\label{eq:non-ren-v}
\frac{Z_1^{-1} b_1^{-1} b_0^{-1}}{Z_0^{-1} b_1^{-2}} =
\frac{Z_0 b_1}{Z_1 b_0} = 1
\end{equation}
which means that the integration shells should actually be
non-spherical and have radii related by
\begin{equation}
b_0 = \frac{Z_0}{Z_1} b_1
\end{equation}
However $Z_0/Z_1=1+O(1/N)$, therefore the deviation from spherical
symmetry is small and is not important to the order of $1/N$ we are
considering.  Thus we can compute the loop integral, assuming that the 
integration region is spherically symmetric.

\begin{figure}[h]
\includegraphics[width=5cm]{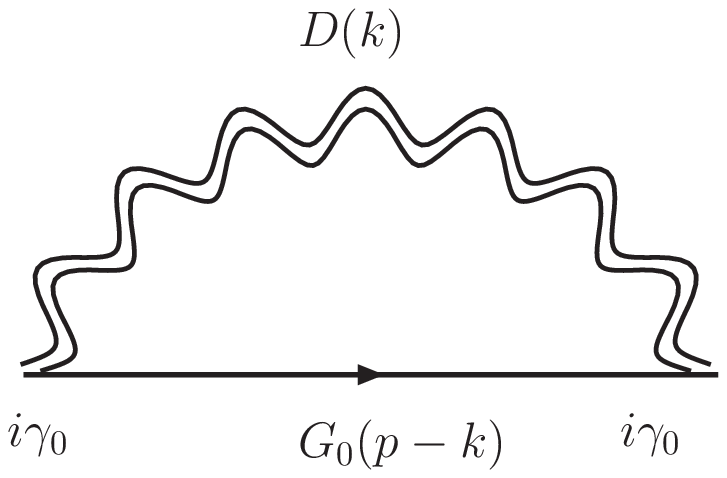}
\caption{\label{fig:SelfEnergy} One-loop self-energy correction to 
the fermion propagator.}
\end{figure}
The  one-loop
self-energy correction to the inverse fermion propagator was
computed in Ref.~\onlinecite{Son}.  In our notation, it corresponds to
\begin{align}
Z_0^{-1} &= 1 - \Delta Z_0 =  1 
  + \frac4{N\pi^2} [I_0(\lambda) - I_1(\lambda)]
     \ln\left [\frac{\Lambda_0}{\Lambda_1}\right ]\label{Z_0}\\
  Z_1^{-1} &= 1 - \Delta Z_1 =  1 
  + \frac4{N\pi^2} I_0(\lambda) \ln\left [\frac{\Lambda_0}{\Lambda_1}\right ]
  \label{Z_1}
\end{align}
The correction to the fermion propagator is suppressed by $1/N$, which is
another reason to work in the large $N$ limit.  Without this small parameter
the diagram of Fig.~(\ref{fig:SelfEnergy}) would be of order one.


To see how the wavefunction renormalization affects the running of the
four-fermi coupling, we write down a generic quartic term in the action 
\begin{equation}
  u\int\!dt\,d^2x \left ( \sum_{a=1}^N {\bar \psi}_a \Gamma \psi_a \right )^2,
\end{equation}
Rescaling of the momenta and the fields as discussed above leads to a
renormalization of the coupling constant $u$ according to
\begin{equation}
\label{eq:CombRes}
u \rightarrow {u'} = 
b_1^{-2} b_0^{-1} ( u + \Delta u ) \left (Z_0^{-1/2} b_1^{-1}\right)^{-4} = 
b_1 \left ( u + \Delta u + u (\Delta Z_0 + \Delta Z_1)\right ) + O(u^2/N)
\end{equation}
where we have also included the contribution $\Delta u$ from the
vertex renormalization, and where $\Delta Z_0 + \Delta Z_1$ comes from
wavefunction renormalization.

From Eqs.~(\ref{Z_0}) and (\ref{Z_1}) we find
\begin{equation}\label{Mwf}
  {\bf M}_{wf}(\lambda) = - \frac 4{N\pi^2}[2I_0(\lambda)-I_1(\lambda)]
\end{equation}

\section{Results and Discussion}
Combining Eqs.~(\ref{Mv}) and (\ref{Mwf}), we find
\begin{equation}
{\bf M}(\lambda) =
\frac{4}{N\pi^2} I_1(\lambda)
\left( \begin{array}{cccccc}
3 & -4 & 0 & 0 & 0 & 0 \\
-2 & 1 & 0 & 0 & 0 & 0\\
0 & 0 & -1 & 0 & 0 & 0\\
0 & 0 & 0 & -1 & 0 & 0\\
0 & 0 & 0 & 0 & 1 & -2\\
0 & 0 & 0 & 0 & -4 & 3
\end{array} \right),
\end{equation}
Notice that the terms containing $I_0(\lambda)$ cancel out, as anticipated, 
so the limit $\lambda \rightarrow \infty$ is finite. 
The eigenvalues yield the anomalous dimensions: 
$\gamma_a = -\frac{8}{N\pi^2} [5,-1,-1,-1,-1,5]$.  The two operators with
lowest anomalous dimensions, $-40/N\pi^2$, are
\begin{equation}\label{operators}
  2(\bar\psi_a\psi_a)^2 - (\bar\psi_a\gamma_0\gamma_i\psi_a)^2, \qquad
  2(\bar\psi_a\gamma_5\psi_a)^2 - (\bar\psi_a\gamma_i\psi_a)^2
\end{equation}

Including the vertices of the form 
$\gamma^3 \Gamma_j \otimes \gamma^3 \Gamma_j$, that were mentioned in the 
introduction, one can repeat the calculation and obtain as a result 
a copy of the above matrix. This is simply a consequence of 
$\{\gamma^3,\gamma^\mu \} = 0$. There is no mixing between these new 
vertices and the ones considered in our calculation. The list of the most 
relevant operators may then be extended to include
\begin{equation}\label{operators2}
  2(\bar\psi_a\gamma^3\psi_a)^2 - (\bar\psi_a\gamma^3\gamma_0\gamma_i\psi_a)^2,
  \qquad
  2(\bar\psi_a\gamma^3\gamma_5\psi_a)^2 - (\bar\psi_a\gamma^3\gamma_i\psi_a)^2
\end{equation}
whose the anomalous dimensions are also $-40/N\pi^2$.

In the large $N$ limit, the anomalous dimensions are small, hence all
four-fermi operators have dimensions close to 4 and are irrelevant.
However, as one decreases $N$ the dimensions of some operators
decrease as well.  Naively, at
\begin{equation}
  N< N_{\rm crit} = \frac{40}{\pi^2} \approx 4.05
\end{equation}
the operators in~(\ref{operators}) and~(\ref{operators2}) would have 
dimensions less than 3, and become relevant.  Of course, this is only
an extrapolation of our leading-order result to finite $N$.  Nevertheless,
it is quite interesting that $N_{\rm crit}$ is close to the real-world
value of $N=4$. (For comparison, for the Lorentz-invariant Thirring model
the critical number of four-component Dirac fermions is quoted as 
$6.6(1)$~\cite{Hands}, to be compared with $N_{\text{crit}}/2 \approx 2$.) 
Due to the limitation of our calculations, we cannot determine whether 
the exact $N_{\rm crit}$ is smaller or larger than 4.

\begin{figure}[h]
\includegraphics[width=11cm]{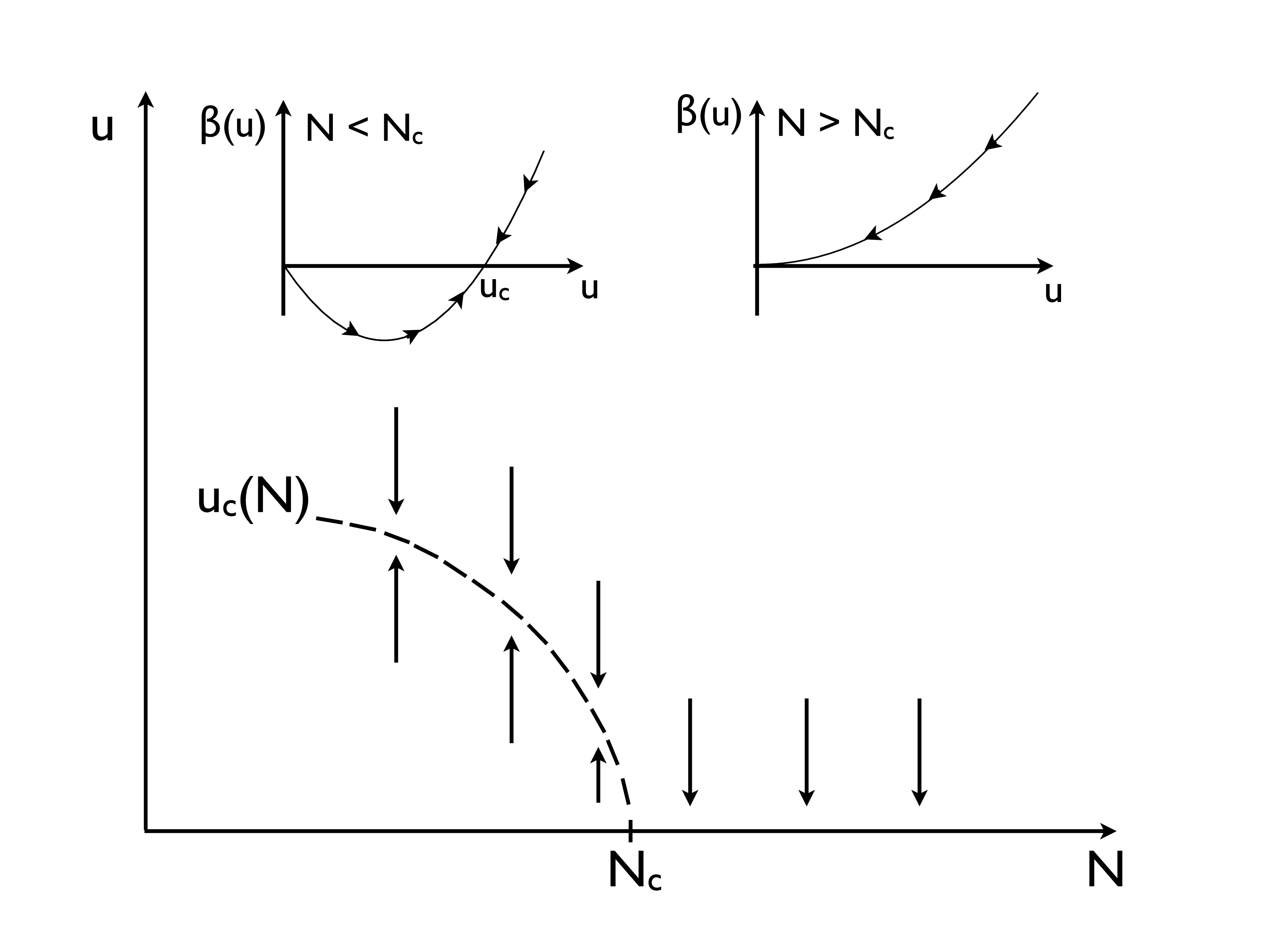}
\caption{\label{fig:PhaseDiagram} Phase diagram and beta functions (Inset), 
in one particular scenario.}
\end{figure}

If $N_{\rm crit}<4$, then the physics at $N=4$ and infinite coupling
is governed by the strong-coupling fixed point discussed in
Ref.~\onlinecite{Son}.  If $N_{\rm crit}>4$, then there are two
further possibilities for $N=4$, $\lambda=\infty$.  It may turn out
that the system develops a gap, and becomes an insulator.  In this case
one expects a bifermion operator to have a nonzero expectation value,
breaking a discrete symmetry.  Unfortunately, simply from the form of the
operators~(\ref{operators}) and (\ref{operators2}) it is not possible
to conclude, with definiteness, which of the discrete symmetries will
be spontaneously broken.
Another possibility is that the coupling flows into a new stable fixed point
(a similar situation was discussed in Ref.~\onlinecite{Herbut}.)
In this case the system remains gapless, but with a new dynamic critical
behavior.  This case is illustrated in Fig.~\ref{fig:PhaseDiagram},
where we also show the $\beta$ function of the coupling $u$.

In real graphene, on the other hand, $\lambda$ is large but not
infinite, so in the above discussion we should take into account that
the anomalous dimensions are functions of both $N$ and $\lambda$. In real 
graphene one has
\begin{equation}
\lambda = \frac{e^2 N}{32 \epsilon_0 \hbar v}
\end{equation}
where  $N=4$ and $v=10^6~\textrm m/\textrm s$, so that the anomalous 
dimension of the operators~(\ref{operators}) and~(\ref{operators2}) is 
$\gamma\approx-0.72$.  Thus, all four-fermi operators are irrelevant, 
at least to the order of $1/N$ we
are considering.  It is conceivable, however, 
that higher-order corrections will
push $\gamma$ to be below $-1$.  On the other hand, for graphene on a
SiO$_2$ substrate with dielelectric constant $\epsilon=5.5$, the
coupling $\lambda$ is reduced by a factor of $2/(1+\epsilon)$, and
$\gamma\approx-0.45$, substantially above $-1$.  
In this case, one can conclude that the
four-fermi interactions are irrelevant (and will become more
irrelevant as the fermion velocity $v$ increases in the infrared).

\section{Conclusion}

We have studied the RG flow of various four-Fermi couplings $u$ in the
low-energy theory of graphene, generalized to include $N$ different
fermion flavors. We computed the anomalous dimensions of the various
couplings to the first nontrivial order in $1/N$.  In the limit of
infinitely strong Coulomb interaction, the operators with lowest
dimensions become relevant for $N<N_{\rm crit}$, where $N_{\rm crit}$
is estimated to be 4.  In real graphene with $N=4$ but at finite
Coulomb coupling, the four-fermi interactions are irrelevant, at least
to the leading nontrivial order in $1/N$.  

It would be interesting to further investigate the phase diagram of
our model.  One can try to push the calculations to another order in
$1/N$.  In addition, one should try to perform numerical simulations
of the model.  The most interesting values of $N$ where nontrivial
phases may exist, as indicated by our calculations, are lower values
like $N=2$ and $N=4$.

\acknowledgments

This work is supported, in part, by DOE grant No.\ DE-FG02-00ER41132.

\smallskip

\emph{Note added:} After this work was completed, the authors learned
of Ref.~\onlinecite{Aleiner}, in which the running of four-fermi
interactions is also considered.  Our lowest anomalous dimension
coincides with the value found in Ref.~\onlinecite{Aleiner}, but our
values of other anomalous dimensions disagree with
Ref.~\onlinecite{Aleiner}.  We thank Igor Aleiner for bringing
Ref.~\onlinecite{Aleiner} to our attention.
 

\end{document}